\documentclass{Interspeech2024}

\usepackage{xcolor}

\usepackage{amsmath}

\DeclareMathOperator*{\argmin}{arg\,min}
\usepackage{multirow}
\usepackage{threeparttable}

\makeatletter
\def\bstctlcite{\@ifnextchar[{\@bstctlcite}{\@bstctlcite[@auxout]}}
\def\@bstctlcite[#1]#2{\@bsphack
  \@for\@citeb:=#2\do{%
    \edef\@citeb{\expandafter\@firstofone\@citeb}%
    \if@filesw\immediate\write\csname #1\endcsname{\string\citation{\@citeb}}\fi}%
  \@esphack}
\makeatother 

\interspeechcameraready

\title{Empowering Whisper as a Joint Multi-Talker and Target-Talker\\Speech Recognition System}

\name{Lingwei Meng, 
Jiawen Kang, 
Yuejiao Wang,
Zengrui Jin,
Xixin Wu,
Xunying Liu,
Helen Meng}

\address{
  The Chinese University of Hong Kong, Hong Kong SAR, China
}
\email{\{lmeng, jwkang, wangy, zrjin, wuxx, xyliu, hmmeng\}@se.cuhk.edu.hk}


\begin{document}
\bstctlcite{IEEEexample:BSTcontrol}

\maketitle

\begin{abstract}
Multi-talker speech recognition and target-talker speech recognition, both involve transcription in multi-talker contexts, remain significant challenges. However, existing methods rarely attempt to simultaneously address both tasks. In this study, we propose a pioneering approach to empower Whisper, which is a speech foundation model, to tackle joint multi-talker and target-talker speech recognition tasks. Specifically, (i) we freeze Whisper and plug a Sidecar separator into its encoder to separate mixed embedding for multiple talkers; (ii) a Target Talker Identifier is introduced to identify the embedding flow of the target talker on the fly, requiring only three-second enrollment speech as a cue; (iii) soft prompt tuning for decoder is explored for better task adaptation. Our method outperforms previous methods on two- and three-talker LibriMix and LibriSpeechMix datasets for both tasks, and delivers acceptable zero-shot performance on multi-talker ASR on AishellMix Mandarin dataset.\footnote{The code is available at https://github.com/LingweiMeng/Whisper-Sidecar}

\end{abstract}

\section{Introduction}
Driven by the rapid development of deep learning along with the availability of large-scale data, automatic speech recognition (ASR) has achieved significant progress in recent years \cite{whisper}. However, speech recognition in the multi-talker scenarios, where overlapping may exist, remains challenging and has attracted much attention.

To tackle the multi-talker speech recognition problem, various approaches have been explored. 
Conventional cascaded systems employ a speech separation module as a front-end to separate mixed speech signals, which are then fed into a single-talker ASR system for transcription \cite{settle2018cascade, li2021cascade}. 
However, these systems usually show limited performance due to the mismatch of their optimization objectives, and may need further joint training \cite{settle2018cascade}. 
Recently, end-to-end models have garnered interest owing to their outstanding performance. 
One primary challenge in training an end-to-end multi-talker ASR system is to associate prediction with the corresponding target labels for loss calculation \cite{jiawen_icassp}. 
Consequently, techniques such as Permutation Invariant Training (PIT) \cite{seki2018pit,wangyou_pit, chang2019pit,chang2020pit,zili_pit_tsr}, Heuristic Error Assignment Training (HEAT) \cite{tripathi2020surt, lu2021surt,surt2}, and Serialized Output Training (SOT) \cite{kanda2020serialized, kanda22b_tsot, chenda} have emerged.
Although these methods have yielded impressive results, they often necessitate training from scratch or performing full fine-tuning on pre-trained models, which 
does not fully capitalize on the existing advancements developed for standard single-talker ASR. 
Enlightened by findings that the ASR encoder captures more acoustic information in its lower layers and more linguistic information in the upper layers \cite{pasad2021layer,shim2021layer,wavlm}, a recent study advocates for the use of a Conv-TasNet-like \cite{luo2019conv} Sidecar separator to tackle multi-talker speech recognition, without distorting the parameters of a well-trained 
single-talker ASR model \cite{sidecar_icassp, sidecar_interspeech}. 

Target-talker ASR, which aims to efficiently recognize speech of a target talker under a multi-talker scenario, also holds substantial practical value. End-to-end approaches have been investigated and achieved substantial progress \cite{zhang23w_interspeech, moriya23_tsasr}. 
However, these methods typically necessitate an external \cite{zili_pit_tsr, ma2023whisper-ts-asr, masumura23joint} or internal \cite{conformer-ts-asr} module to derive the speaker embedding from the enrollment speech of target-talker, consequently increasing the model's computational burden. Moreover, they typically only output the transcripts of an assigned target talker, neglecting the speech of other talkers. This limitation hinders their applicability in situations where users may also be interested in obtaining the transcriptions of non-target talkers.
Although speaker-attributed ASR can transcribe multiple speakers in a speaker-aware manner, it typically necessitates the speaker embeddings of all involved individuals \cite{kanda21b_sasot}.
As far as we know, \cite{masumura23joint} is the only study attempting to address joint multi-talker and target-talker ASR; however, it still requires an external speaker embedding extractor.

Nowadays, speech foundation models have emerged as a versatile solution for diverse speech tasks \cite{baevski2020wav2vec, hubert, wavlm}. As an representative in this domain, Whisper \cite{whisper} has demonstrated its potential across various tasks beyond ASR \cite{whisper-at-2023,peng23d_prompt_whisper} which motivated us to further extend Whisper’s capabilities in tackling multi-
talker and target-talker speech recognition challenges.

In this study, we empower Whisper as a joint multi-talker and target-talker system in a parameter-efficient style. Specifically, we freeze the weights of Whisper and incorporate a Sidecar separator into its encoder to endow it with multi-talker speech recognition capabilities. A Target Talker Identifier (TTI) module is introduced to distinguish the target speaker's embedding branch on the fly, requiring only three seconds of the target talker's enrollment speech as a cue. Moreover, soft prompt tuning \cite{lester2021prompt} for Whisper decoder is adopted to further adapt to the tasks. Our major contributions are threefold:

\vspace{-0.5mm}
\begin{itemize}
\setlength\itemsep{-0.2mm}
    \item We propose a pioneering framework to jointly transcribing multi-talker speech while highlighting the target talker's speech, without employing any speaker embedding extractor. 
    \item Leveraging the frozen Whisper as the foundation model, our framework only involves limited trainable parameters, making it a parameter-efficient and loosely-coupled system.
    \item Extensive experiments reveal that the proposed approach achieves leading performance on two- and three-talker LibriMix and LibriSpeechMix datasets (English) on both tasks, and attains satisfactory zero-shot multi-talker ASR performance on AishellMix (Mandarin).
\end{itemize}

\vspace{-5mm}

\begin{figure*}[htbp]
\begin{center}
\includegraphics[width=0.96\textwidth]{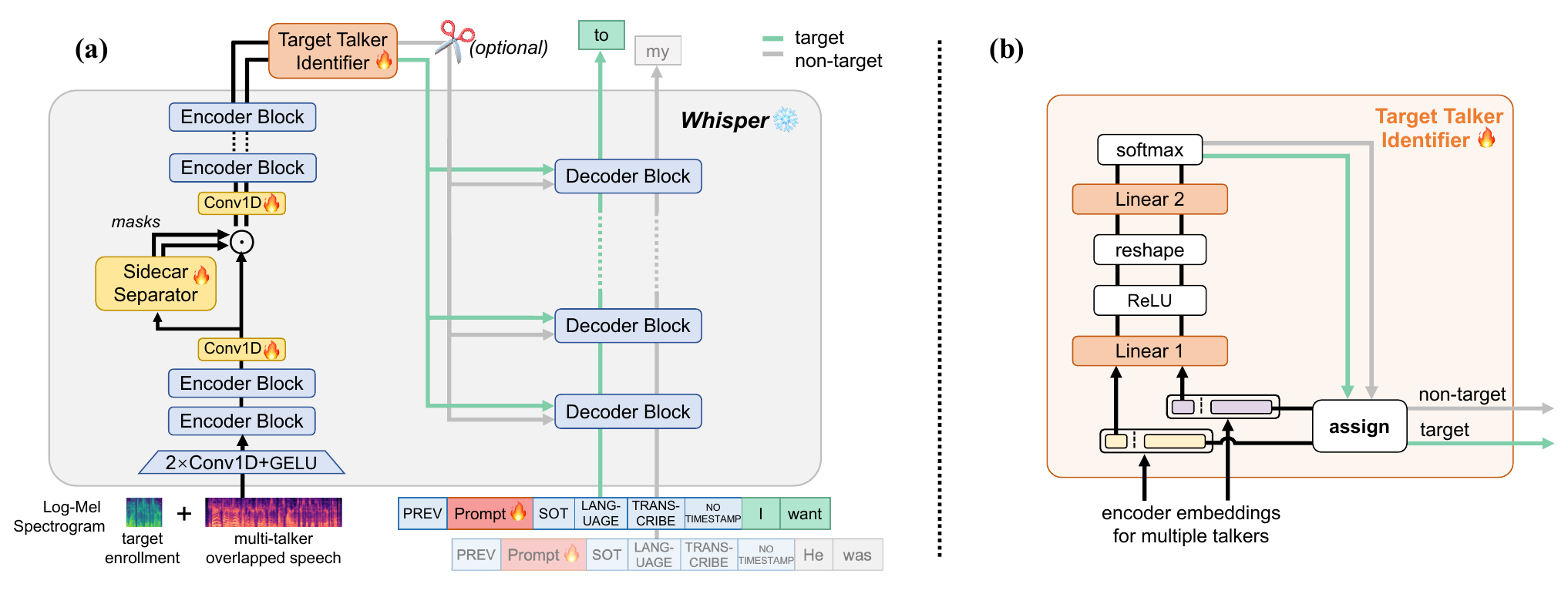}
\end{center}
\vspace{-6mm}
\caption{Take two-talker scenario as an example, the proposed system (a) take the concatenation of target enrollment speech and multi-talker speech as input. The embedding is separated by Sidecar separator. Target Talker Identifier (b) processes the prefix segments of encoder embeddings to identify the target talker branch. Optionally, non-target branchs can be discarded to accelerate inference.}
\vspace{-4mm}
\label{fig:framework}
\end{figure*}


\section{Methods}

The proposed method consists of four main components — Whisper serving as the foundation model, a Sidecar separator to separate mixed embedding for multiple talkers, a Target Talker Identifier to identify the embedding flow of the target talker, and a soft prompt embedding to facilitate task adaptation.
\vspace{-1mm}
\subsection{Whisper as the Speech Foundation Model}
Whisper is a speech recognition model featuring an attention-based encoder-decoder structure, which has been trained on massive amounts of web-scale labeled speech data \cite{whisper}. Nowadays, it is increasingly being utilized as a speech foundation model even beyond speech recognition \cite{ma2023whisper-ts-asr, whisper-at-2023, peng23d_prompt_whisper}.
In this study, we are inspired to extend its capability to handling joint multi-talker and target-talker ASR tasks.

Whisper takes log-Mel spectrogram as input, followed by transformer encoder and decoder modules to decode the output tokens in an auto-regressive manner.
Different from other ASR models, Whisper adopts several special tokens as the prefix of input sequences for decoder to specify tasks and condition information.
By default, the prefix tokens are \texttt{"<|PREV|>}, \textit{text prompt}, \texttt{<|SOT|>}, \texttt{<|LANGUAGE|>}, \texttt{<|TRANSCRIBE|>}, \texttt{<|NO\_TIMESTAMP|>"}, where \texttt{<|PREV|>} and \textit{text prompt} are optional.



\vspace{-1mm}
\subsection{Empowering Whisper as a Multi-Talker ASR System}
Recently, the Sidecar separator (SS) has been introduced as a parameter-efficient module to convert a well-trained single-talker ASR model into a multi-talker one \cite{sidecar_icassp}. 
In this work, we incorporated the Sidecar separator with Whisper to harness its capability acquired from extensive training data.

The Sidecar separator is a temporal convolutional network inserted between the early layers of the ASR encoders.
It consists of stacked 1-D dilated convolutional blocks inspired by Conv-TasNet \cite{luo2019conv}.
As the shallower layers of the ASR encoder are believed to encode more acoustic information rather than the linguistic ones\cite{shim2021layer, sidecar_icassp}, the Sidecar separator is able to separate the mixed representation with talker-related masks, producing disentangled representation of different speakers.

As depicted in Figure \ref{fig:framework}, the Sidecar separator accompanied by two 1-D convolutional layers is positioned after the second encoder block.
Talker-dependent masks are generated, which are element-wisely multiplied with the mixed embedding, yielding separated embeddings of each talker. 
The subsequent encoder blocks and decoder process these branches, ultimately transcribing the corresponding text for each talker.



\vspace{-1mm}
\subsection{Target Talker Identifier}
\label{sec:TTI}
We introduce the Target Talker Identifier (TTI) module, which equips the system with the capability for target-talker ASR. 

During the forward process, as illustrated in Figure \ref{fig:framework} (b), the encoder-output embeddings corresponding to different talkers will be segmented into two distinct segments: 
a prefix segment aligning with the length of three-second enrollment speech, and a main segment that corresponds to the duration of the multi-talker speech. 
The prefix segments are then fed into the TTI module, which determines the branch associated with the target talker, while only the main embedding segments are sent to the Whisper decoder for transcription.


Specifically, the prefix segments hold a tensor shape of $(B\times S, \textit{150}, C)$, where $B$ denotes batch size, $S$ denotes the number of talkers, $C$ denotes the number of channels, and $\textit{150}$ denotes the number of time frames. Given that each time frame spans a duration of 20 ms, $\textit{150}$ coincides with the three-second duration of the enrollment speech. As shown in Figure 1, within the TTI module, the prefix segments traverse a linear layer followed by the ReLU activation function, yielding a tensor with a shape of $(B\times S, \textit{150}, \textit{1})$. Upon squeezing and reshaping, the tensor proceeds through another linear layer and the softmax function to produce the probability $(B, S)$ of each branch being the target talker.

Consequently, the target talker branch is efficiently determined on the fly, introducing minimal computational overhead. Underpinned by the superior performance of the separation module, TTI can be considered as performing target-talker activity detection, which is a more economical task compared with methods necessitating speaker embedding extraction.

\vspace{-1mm}
\subsection{Soft Prompt Tuning}
The original Whisper model allows for the inclusion of text prompt tokens as prefix to the decoder's input sequences, conditioned on which the model yields improved ASR performance on ambiguous audios \cite{whisper}. In this context, by exploiting this inherent characteristic of Whisper combined with soft prompt tuning technique \cite{lester2021prompt}, we aim to adapt the model more efficiently to multi-talker and target-talker ASR tasks.

Specifically, as shown in Figure \ref{fig:framework} (a), we insert a learnable embedding as soft prompt between \texttt{<|PREV|>} and \texttt{<|SOT|>} tokens where hard prompt tokens were originally specified. Note that we mask the position of the soft prompt when calculating the training loss, since the model does not require learning to generate them. The soft prompt embedding will be updated as the model learns to transcribe the multi-talker speech.

\vspace{-1.5mm}
\subsection{Training objectives}
\vspace{-0.5mm}
At each training step, there's an 80\% probability of undertaking multi-talker ASR training, while a 20\% probability for appending a three-second enrollment speech for joint multi-talker and target-talker ASR training.
Both ASR loss and TTI's cross-entropy loss necessitate a permutation assignment for speaker order to address the label ambiguity issue \cite{yu_pit}. In this study, the permutation is determined by Permutation Invariant Training (PIT) based on ASR loss, and is then assigned for TTI's cross-entropy loss calculation. The permutation is derived as:
\setlength{\abovedisplayskip}{3pt}
\setlength{\belowdisplayskip}{3pt}
\begin{align}
\hat{\pi} & = \argmin_{\pi\in\mathcal{P}}\sum_{s=1}^{S}\,\text{Loss}_\text{ASR}(Y^s, R^{\pi(s)})
\end{align}
\noindent
where $\mathcal{P}$ denotes the set of all permutations  on $=\{1, ... , S\}$, $\pi(s)$ denotes the $s$-th element in a permutation $\pi$, $Y^s$ is the predicted token sequences of the $s$-th branch, and $R$ is the reference labels for $S$ talkers.
At last, the final objective function is the sum of PIT-ASR loss and corresponding TTI loss multiplied by a coefficient $\lambda$. Therefore we have,
\begin{align}
&\mathcal{L}  = \mathcal{L}_\text{ASR} + \lambda\,\mathcal{L}_\text{TTI} \\
&\mathcal{L}_\text{ASR}  = \sum_{s}\,\text{Loss}_\text{ASR}(Y^s, R^{\hat{\pi}(s)}) \\ 
&\mathcal{L}_\text{TTI} = \text{Loss}_\text{CE}(Z, D^{\hat{\pi}})
\end{align}

\noindent
where $Z$ is the probability of each branch is of the target talker, and $D^{\hat{\pi}}$ is the ground truth after permutation.

Considering that the original Whisper was not trained using CTC loss, we refrain from employing an additional CTC loss for early permutation assignment as done in \cite{seki2018pit,wangyou_pit}. 

\vspace{-1.5mm}

\section{Experimental Setup}
\vspace{-1mm}
\subsection{Datasets}
\vspace{-0.5mm}
The experiments are conducted on three multi-talker public datasets, namely LibriMix \cite{cosentino2020librimix} and LibriSpeechMix \cite{kanda2020serialized} in English, and Aishell1Mix\footnote{https://github.com/huangzj421/Aishell1Mix} in Mandarin. Audio exceeding Whisper's maximum handling duration of 30 seconds are time-stretched to conform to this limit. For target-talker ASR on LibriMix and LibriSpeechMix, we randomly trim three-second clips from LibriSpeech as enrollment speech for each talker.


\noindent\textbf{LibriMix}. The dataset simulates audio mixtures in a left-aligned manner, involving two or three speakers from the LibriSpeech-clean corpus. Thus, the shorter source speech is entirely overlapped with the longer one from the start, presenting significant challenge in separating overlaps. We focus on its two-speaker-mixed and three-speaker-mixed clean subset, denoted as Libri2Mix and Libri3Mix in the following.

\noindent\textbf{LibriSpeechMix}. The utterances are simulated from LibriSpeech, comprising mixtures from two or three speakers. Unlike LibriMix, the delay time for each speaker is randomly sampled, resulting in partially overlapped mixtures. Since only official dev and test sets are released, we created our training set from the 960-hour LibriSpeech following the same protocol as in \cite{kanda2020serialized}, except that the mixtures are kept under 30 seconds.

\noindent\textbf{Aishell1Mix}. It is a  Mandarin multi-talker speech dataset, source from Aishell1 corpus. It simulate mixtures with a same protocol of the LibriMix. We focus on its two-speaker-clean subset for analysis, denoted as AishellMix in the following.

\vspace{-1.5mm}

\subsection{Model Settings and Evaluation Metrics}
\vspace{-0.5mm}
Throughout this study, we employ Whisper-small, -medium, and -large-v3 as the foundation models, respectively. We freeze these models and only train the Sidecar separator, Target Talker Identifier, and soft prompt embedding. The number of trainable parameters for systems using different foundation models and for various numbers of talkers are listed in Table \ref{tab:param}.

The Conv-TasNet-like Sidecar separator \cite{sidecar_icassp} comprises a series of \textit{K} temporal convolutional blocks with dilation rates ranging from 1 to $2^{\textit{K} - 1}$, with each block repeats up to \textit{R} times. Consistent with the protocol in \cite{sidecar_icassp,sidecar_interspeech}, we use \textit{K} = 8 and \textit{R} = 3 and plug it between the second and third encoder blocks. The length for soft prompt embeddings are investigated through ablation experiments (Section \ref{sec:ablation}). As a result, we establish a length of $\textit{4}$, which gives the best performance.

For systems with the TTI module, at each training step, there's an 80\% probability of undertaking multi-talker ASR training, while a 20\% probability for joint multi-talker and target-talker ASR training. We set the coefficient of TTI loss $\lambda$ to 0.01. The systems are trained and evaluated on two- and three-talker subsets of LibriMix and LibriSpeechMix, respectively. Each training session lasts for a maximum of 200k steps on 8 NVIDIA V100 GPUs with a total batch size of 16, employing AdamW optimizer with an initial learning rate of 2e-4 that decreases linearly to 1e-4.

Permutations with minimum errors are used to compute word error rate (WER) or character error rate (CER) for multi-talker ASR as in prior studies \cite{kanda22b_tsot, sidecar_icassp}. For target-talker ASR, we use standard WER for evaluation. 
Both the model's predictions and the references are normalized following \cite{whisper}.
\vspace{-1mm}

\begin{table}[!h]
\vspace{-2mm}
  \caption{The amount of trainable parameters, with numbers in parentheses indicating their proportion in the total parameters.}
\vspace{-3mm}
  \label{tab:param}
  \centering 
  \setlength{\tabcolsep}{1mm}
  {
  \begin{tabular}{lcccc}

    \toprule
    \textbf{Foundation Model} & 2-speaker & 3-speaker     \\
    \midrule
    Whisper-small   &   8.69 M (3.47\%) &  8.79 M (3.51\%)     \\
    Whisper-medium    & 13.16 M (1.69\%)   & 13.29 M (1.71\%)   \\
    Whisper-large    &  18.41 M (1.18\%) & 18.58 M (1.19\%)     \\

    \bottomrule
    \end{tabular}
    }
\vspace{-3mm}
\end{table}





\vspace{-1.5mm}

\section{Results and Discussions}
\subsection{Multi-Talker ASR Results}
We compared the performance of various systems for multi-talker ASR on the two- and three-talker LibriMix and LibriSpeechMix test sets, as shown in Table \ref{tab:mtasr}. Empowered by Sidecar separator (SS), TTI, and soft prompt, our systems (f)-(k) consistently illustrated improved performance across both the two- and three-speaker subsets. Even with Whisper-small-SS-TTI (g), owing to Whisper's extensive pre-training, our method has already surpassed the original Sidecar scheme\cite{sidecar_interspeech}. As the size of the Whisper model increases, we observed a steady improvement in performance, which aligns with our expectations. Ultimately, our systems outperformed previous approaches across all datasets except for LibriSpeechMix-3spk, demonstrating the superiority of the proposed approach.

Interestingly, we find that systems with the TTI module (g) (i) (k) outperform their counterparts without TTI (f) (h) (j) in multi-talker ASR task, even though the TTI is specifically designed to support target-talker ASR. This suggests that the training objective of learning to distinguish the target talker can benefit the Sidecar's ability to separate embeddings, thereby facilitating the task of multi-talker ASR.

\vspace{-1.5mm}
\subsection{Target-Talker ASR Results}
\vspace{-1mm}
We evaluate target-talker ASR performance on two- and three-speaker subsets of LibriMix and LbriSpeechMix, as illustrated in Table \ref{tab:ttasr}.\footnote{We did not include results reported in \cite{conformer-ts-asr}, which delivers better performance but undergoes about ten times training efforts as ours.} Our systems outperform previous state-of-the-art method on LibriMix dataset by a large margin, and we are the first to perform target-talker ASR task on Libri3Mix.
To guarantee a fair comparison, we further trained three additional systems with limited training data to ensure consistency with the data used in \cite{zili_pit_tsr, ma2023whisper-ts-asr}. These systems are denoted as "-limited".
The results demonstrate that our method still outperforms \cite{zili_pit_tsr, ma2023whisper-ts-asr} though under this restriction.

For LibriSpeechMix, the speech signals are partially overlapped, which means the target talker’s speech can incur a considerable delay before it commences, resulting in a substantial time interval away between it and the enrollment speech. 
Nevertheless, despite the existence of delays, our systems still demonstrate good performance on the LibriSpeechMix dataset, validating the effectiveness of the proposed method.  

\vspace{-1.5mm}
\subsection{Zero-Shot Multi-Lingual Evaluation}
\vspace{-1mm}
We investigated whether the multi-lingual characteristics of Whisper are retained after fine-tuning it on an English multi-talker dataset. Specifically, we conducted evaluations for systems (g) (i) (k) listed in Table 1 using the two-speaker AishellMix Mandarin dataset, which is the first time to be used on multi-talker ASR task.
The evaluations are performed using two schemes: zero-shot and one-batch-tuning. Zero-shot refers to directly evaluating the system on the AishellMix, while one-batch-tuning implies conducting an additional training epoch on the AishellMix training set prior to the evaluation.

As Table 5 illustrates, the medium and large models demonstrated acceptable CER performance even under zero-shot conditions. With just one batch tuning, these models exhibited satisfactory results. This suggests that our method largely maintains the inherent multilingual capabilities of Whisper.

\vspace{-1.5mm}
\subsection{Ablation Study}
\vspace{-0.5mm}
\label{sec:ablation}
We investigated the optimal prompt length by examining the multi-talker ASR performance on the Libri2Mix dataset with Whisper-medium and Whisper-large models. As shown in Table \ref{tab:ablation},  a soft prompt of length 4 yields the best performance. However, as the soft prompt length increases to 16, the systems see a decline in performance. This may be due to overly long sequence sequences, making the model difficult to optimize, given the original Whisper model is frozen.
\vspace{-1.5mm}
\subsection{Limitations and Future Work}
\vspace{-1mm}
This study has several limitations. Firstly, our method relies on PIT which requires pre-defining of the maximum number of speakers. Future efforts will integrate SOT \cite{kanda2020serialized} or HEAT \cite{lu2021surt} to address this issue and reduce training costs. 
Secondly, when the target talker's speech undergoes excessive delay, there could be potential degradation in the target-talker ASR's performance. We anticipate future work to enable the TTI module synthesize information across the entire utterance duration rather than only the three-second enrollment speech.

\begin{table}[!t]
\begin{flushleft}

  \footnotesize

  \caption{Multi-talker ASR on the test sets of LibriMix and LibriSpeechMix. Evaluated by WER (\%). “SS” denotes “Sidecar Separator”, “TTI” denotes “Target Talker Identifier”.}
  \label{tab:mtasr}
  \centering 
\vspace{-3mm}
  {
  \begin{tabular}{lccccccccc}

    \toprule
 & \multicolumn{2}{c}{\textbf{LibriMix}} & \multicolumn{2}{c}{\textbf{LibriSpeechMix}} \\
 \cmidrule(r){2-3} \cmidrule(r){4-5}
    \textbf{System} & 2spk & 3spk   &    2spk & 3spk     \\
  \midrule
   (a) WawLM Base+ PIT \cite{zili_pit_tsr} &  18.45 &-& - &-\\
   (b) C-HuBERT-Large \cite{cocktail_hubert}  &  7.80 &-& - &-  \\
    (c)    SURT \cite{lu2021surt} &  -&-& 7.20  &-\\
    (d) SOT-Conformer \cite{kanda21b_sasot}&  - &-&4.90\textsuperscript{\dag}  &\textbf{6.20}\textsuperscript{\dag}   \\ 
      (e) D2V-Sidecar-DB \cite{sidecar_interspeech} & 9.69 & 33.91  & 7.49 & 11.94\\
    \midrule
  (f) Whisper-small-SS &  10.04 &   29.20  &  5.27 & 9.85    \\
   (g) Whisper-small-SS-TTI  & 9.39   & 26.76 & 5.18 & 8.61    \\
   
  (h) Whisper-medium-SS   & 6.95  & 22.58  &  4.32 &  7.80 \\
  (i) Whisper-medium-SS-TTI   & 6.56 & 21.47  & 4.01 & 7.50  \\
   
  (j) Whisper-large-SS    &  4.98 &  17.55 &  3.81 & 7.13  \\
  (k) Whisper-large-SS-TTI    &  \textbf{4.66} & \textbf{16.79} & \textbf{3.43} & 6.80 \\
   
\bottomrule
\end{tabular}

}

\begin{tablenotes}
\item\textsuperscript{\dag} with extremely heavier training efforts.
\end{tablenotes}  

\vspace{-5mm}
\end{flushleft}
\end{table}
\renewcommand{\arraystretch}{1}
\begin{table}[!t]
  \footnotesize
  \caption{Target-talker ASR on LibriMix and LibriSpeechMix. Evaluated by WER (\%). "-limited" denotes using the same training data as in \cite{ma2023whisper-ts-asr}. }
  \label{tab:ttasr}
  \centering 
  \vspace{-3mm}
  \setlength{\tabcolsep}{1.4mm}
  {
  \begin{tabular}{lcccc}

    \toprule
    & \multicolumn{2}{c}{\textbf{LibriMix}} & \multicolumn{2}{c}{\textbf{LibriSpeechMix}} \\
    \cmidrule(r){2-3} \cmidrule(r){4-5}
    \textbf{System} & 2spk & 3spk   &    2spk & 3spk     \\
    \midrule
    WavLM-Base$^{+}$-TSE \cite{zili_pit_tsr} & 12.32   &  -   & -  & -    \\
    Whisper-TS-ASR \cite{ma2023whisper-ts-asr} & 11.98   &  -   & -  & -    \\
    \midrule
    Whisper-small-SS-TTI-limited    &  15.75  &  - &  -  &  -   \\
    Whisper-medium-SS-TTI-limited    &   11.39 &  -  & -   & -  \\
    Whisper-large-SS-TTI-limited   &  10.79  &   - &  -  &  - \\
    Whisper-small-SS-TTI    &   11.81 &   30.52  &   8.89 & 15.85    \\
    Whisper-medium-SS-TTI    &  9.14  & 25.75  &  7.58  &  12.4 \\
    Whisper-large-SS-TTI    &  \textbf{7.97}  & \textbf{21.97}  &  \textbf{6.99}  &  \textbf{11.4}  \\

    \bottomrule
    \end{tabular}
    }
\vspace{-1mm}
\end{table}




\begin{table}[!t]
  \footnotesize
  \caption{Zero-shot and one-batch-tuning multi-talker ASR on Aishell1Mix Mandarin dataset. Evaluated by CER (\%).}
  \label{tab:zh}
  \centering 
  \vspace{-3mm}
  \setlength{\tabcolsep}{3.6mm}
  {
  \begin{tabular}{lcccc}

    \toprule
    \textbf{System} & \textbf{zero-shot} & \textbf{one-batch-tuning}     \\
    \midrule
    Whisper-small-SS-TTI    &   55.87 &  28.95     \\
    Whisper-medium-SS-TTI    & 36.28   & 19.83   \\
    Whisper-large-SS-TTI    &   \textbf{28.94} & \textbf{17.81}     \\

    \bottomrule
    \end{tabular}
    }
\vspace{-0.1cm}
\end{table}





\begin{table}[!t]
\footnotesize
  \caption{Ablation study on soft prompt, evaluated by WER (\%).}
  \label{tab:ablation}
  \centering
  \vspace{-3mm}
    \setlength{\tabcolsep}{2.3mm}
    {
  \begin{tabular}{lccccccccc}
\toprule
&\multicolumn{5}{c}{\textbf{Soft Prompt Length}}\\
\cmidrule(r){2-6}
  \textbf{System} & 0 & 2&4 & 8& 16 \\
  \midrule

  Whisper-medium-SS-TTI & 7.21  & 6.82 & \textbf{6.56} & 6.84 &  7.5 \\
  Whisper-large-SS-TTI  &  5.27 & 4.98  & \textbf{4.66}  & 4.74 & 5.43 \\
\bottomrule
\end{tabular}}
\vspace{-0.25cm}
\end{table}

\section{Conclusions}
In this study, we introduce a novel methodology that harnesses Whisper, a speech foundation model, to jointly transcribe multi-talker speech meanwhile highlighting the target talker’s
speech, without employing any speaker embedding extractor.
Specifically, we freeze whisper and insert a Sidecar separator into its encoder to separate mixed embedding for multiple talkers. Subsequently, a Target Talker Identifier module is introduced to identify the embedding flow of the target talker on the fly, requiring only three-second enrollment speech as a cue. The soft prompt tuning is further utilized to facilitate task adaptation. 

Extensive experiments reveal that our approach outperforms previous methods on LibriMix and LibriSpeechMix on both tasks. Moreover, it achieves acceptable zero-shot performance on multi-talker ASR on AishellMix Mandarin dataset.

\clearpage  
\newpage

\section{Acknowledgements}
This research is partially supported by the HKSARG Research Grants Council’s Theme-based Research Grant Scheme (Project No. T45-407/19N) and by the CUHK Stanley Ho Big Data Decision Analytics Research Centre.

\bibliographystyle{IEEEtran}
\bibliography{mybib}

\end{document}